\documentclass[journal]{IEEEtran}
\usepackage{cite}
\usepackage{amsmath} 

\usepackage{subfigure}
\ifCLASSINFOpdf
\usepackage[pdftex]{graphicx}
  \graphicspath{{../pdf/}{../jpeg/}}
  \DeclareGraphicsExtensions{.pdf,.jpeg,.png}
\else
  \usepackage[dvips]{graphicx}
  \graphicspath{{../eps/}}
  \DeclareGraphicsExtensions{.eps}
\fi  
\usepackage{amsmath}
\usepackage{cases}
\usepackage{stfloats}
\usepackage{amsfonts}
\usepackage{subeqnarray}
\usepackage{longtable}
\usepackage{supertabular}
\usepackage{setspace}
\usepackage{multirow}
\usepackage{booktabs}
\usepackage[ruled,linesnumbered]{algorithm2e}
\makeatletter
\newcommand{\nosemic}{\renewcommand{\@endalgocfline}{\relax}}% Drop semi-colon ;
\newcommand{\dosemic}{\renewcommand{\@endalgocfline}{\algocf@endline}}% Reinstate semi-colon ;
\let\oldnl\nl% Store \nl in \oldnl
\newcommand{\nonl}{\renewcommand{\nl}{\let\nl\oldnl}}% Remove line number for one line
\makeatother
\usepackage[export]{adjustbox} 
\usepackage{booktabs}
\usepackage{setspace}
\usepackage{xcolor}
\usepackage{mathrsfs}
\usepackage{amsmath}
\usepackage{array}
\usepackage{amssymb}
\usepackage{amsthm}
\usepackage{microtype}
\usepackage{url}
\usepackage{amsfonts,amssymb}
\usepackage{dsfont}
\usepackage{mathtools}
\usepackage{xcolor,colortbl}
\usepackage{colortbl}
\usepackage{graphicx}
\usepackage{enumitem}
\usepackage[utf8]{inputenc}
\usepackage{booktabs}

\definecolor{Gray}{gray}{0.85}
\definecolor{Whitecolor}{rgb}{1,1,1}
\definecolor{cincinnati-red}{RGB}{190,0,0}
\definecolor{purpple}{RGB}{251,142,50}

\newcolumntype{a}{>{\columncolor{Gray}}c}
\newcolumntype{b}{>{\columncolor{white}}c}
\allowdisplaybreaks
\begin{document}
\setstretch{1.053}
\title{\textls[-25]{A Distributionally Robust Control Strategy for Frequency Safety based on Koopman Operator Described System Model}}
\author{Qianni~Cao,~\IEEEmembership{Student~Member,~IEEE},
Chen~Shen,~\IEEEmembership{Senior~Member,~IEEE}
\thanks{The work is funded by the National Key R\&D Program of China
“Response-driven intelligent enhanced analysis and control for bulk power system stability”(2021YFB2400800) and the science and technology project of SGCC (State Grid Corporation of China): The key technologies for electric internet of things(SGSDDK00WJJS2200092).}
% \thanks{The next few paragraphs should contain
% the authors' current affiliations, including current address and e-mail. For
% example, F. A. Author is with the National Institute of Standards and
% Technology, Boulder, CO 80305 USA (e-mail: author@boulder.nist.gov). }
\thanks{Qianni Cao is with the Electrical Engineering Department, Tsinghua University, Beijing, China (e-mail: cqn20@mails.tsinghua.edu.cn).Chen Shen is with
the Electrical Engineering Department, Tsinghua University, Beijing, China (e-mail: shenchen@mail.tsinghua.edu.cn).}

%\thanks{M. Jia and C. Shen are with the State Key Laboratory of Power Systems, Tsinghua University, 100084 Beijing, China. Y. Wang and G. Hug are with the Power Systems Laboratory, ETH Zurich, 8092 Zurich, Switzerland.}
}
        
%\thanks{This work was supported in part by the Joint Funds of the National Natural Science Foundation of China under Grant U1766206 (Correspondence to Chen Shen).}
%\thanks{M. Jia, C. Shen and Z. Wang are affiliated with the State Key Laboratory of Power Systems, Department of Electrical Engineering, Tsinghua University, Beijing 100084, China (e-mail addresses: jms16@mails.tsinghua.edu.cn, shenchen@mail.tsinghua.edu.cn,
%    wangzhaojian@mail.tsinghua.edu.cn).}% <-this % stops a space
% \thanks{Manuscript received April 19, 2005; revised August 26, 2015.}

%\markboth{Submitted to IEEE Trans. Smart Grid}%
\maketitle

\begin{abstract}
As the proportion of renewable energy and power electronics in the power system increases, modeling frequency dynamics under power deficits becomes more challenging. Although data-driven methods help mitigate these challenges, they are exposed to data noise and training errors, leading to uncertain prediction errors. To address uncertain and limited statistical information of prediction errors, we introduce a distributionally robust data-enabled emergency frequency control (DREFC) framework. It aims to ensure a high probability of frequency safety and allows for adjustable control conservativeness for decision makers. Specifically, DREFC solves a min-max optimization problem to find the optimal control that is robust to distribution of prediction errors within a Wasserstein-distance-based ambiguity set. With an analytical approximation for VaR constraints, we achieve a computationally efficient reformulations. Simulations demonstrate that DREFC ensures frequency safety, low control costs and low computation time.
\end{abstract}
%Although the historical data of renewable generations could be assumed as publicly known
\IEEEpeerreviewmaketitle

% \section*{Nomenclature}
% \addcontentsline{toc}{section}{Nomenclature}

% \subsection*{Indices} 
% \begin{IEEEdescription}[\IEEEusemathlabelsep\IEEEsetlabelwidth{$aaaaaaaa$}]

% \end{IEEEdescription}

\section{Introduction}
\subsection{Motivation}
\IEEEPARstart{A}{s} the proportion of renewable energy increases, the complexity and uncertainty of grid operations will also grow. The variability of operating conditions, uncertainty of disturbances and faults, weak disturbance resistance of power electronic devices, and high flexibility of control make response-driven emergency frequency control increasingly important.

Response-driven emergency frequency control typically employs historical data to model system dynamics and forecast future system trajectories. These forecasts serve as constraints, enabling the design of optimization-based control strategies. However, using historical data to model system dynamics may encounter the following issues: 1) Data acquisition phase: issues such as measurement noise, communication delays, and limited data availability and 2) Dynamics representation phase: potential errors such as incorrect assumptions about the library of basis functions and training inaccuracies. These factors lead to unavoidable and highly uncertain prediction errors in data-driven system dynamics representation. Consequently, using these predictions as constraints in optimal control problems may not yield optimal solutions. Addressing the uncertainty in prediction errors is a critical concern.

Currently, extensive research has been conducted on addressing uncertainty in power systems. Among these, stochastic optimization (SO) and robust optimization (RO) have been widely applied. While SO is well-established for managing uncertainty, it assumes random variables follow a specific distribution, making it less robust for extreme scenarios. RO seeks optimal decisions under worst-case scenarios by setting bounds on uncertain parameters and is widely applied in power systems. However, RO's lack of probabilistic information results in conservative outcomes. Consequently, distributionally robust optimization (DRO), which considers the worst-case PD of uncertain parameters, is gaining attention.

This paper proposes a new distributionally robust emergency frequency control model, which incorporates Value at Risk (VaR) constraints to ensure a high probability of frequency safety. A data-driven GMM-based ambiguity set, explicit and convex, is used to hedge against the uncertain true distribution of frequency prediction error. An approximate method for the VaR constraint is then proposed, allowing it to be expressed analytically. Unlike DRO based on empirical distribution, the proposed method's computational efficiency is not significantly affected by the increase in historical data.

\subsection{Literature Review}

Modern power systems require emergency frequency control (EFC) strategies to manage significant under-frequency events. The traditional second defense line approach, which involves reliable solutions in advance, ensures system stability during rare, severe faults. However, as renewable energy penetration increases, time-domain simulations struggle to handle the complexity and variability of operating conditions, leading to a higher risk of strategy mismatch. Consequently, data-driven methods are gaining traction in the industry. These methods leverage collected data to capture system dynamics, predict future behaviors, and compute optimal control sequences. Data-driven models offer rapid updates based on system responses, addressing the challenges of varying conditions and overcoming limitations of traditional mechanistic modeling. For example, Ref. \cite{zhang2023cnn} proposed to use of a CNN-LSTM-based strategy has improved voltage stability by adaptively evaluating the stability margin. Ref. \cite{Xu2015} proposed a coordinated AC/DC control based on real-time power system response data.

However, data-driven control continues to face challenges, such as noisy data and training bias, which can introduce inaccuracies in predicting future trajectories and undermine effective system stabilization.

A common approach to addressing these challenges is the integration of robust control. For linear systems, Ref. \cite{9468332} proposed a data-driven robust model predictive control based on the Fundamental Lemma for grid-following converters. Additionally, Ref. \cite{10034855} theoretically derives the relationship between the min-max formulation of robust control and regularization problems, applying this method to design data-driven control for grid-following converters. This control strategy is further extended to nonlinear systems in Ref. \cite{10319277}. For nonlinear systems, Ref. \cite{cao2023datadrivenfrequencyloadshedding} introduces a stability margin design for frequency security, based on a system model described by the Koopman operator, which accounts for prediction errors and control deviations in the data-driven model. Ref. \cite{ZHANG2022110114} proposes a robust model predictive control based on the Koopman operator, considering additive model errors and providing a box uncertainty set representation for data-driven additive error uncertainty.

However, robust control can sometimes be overly conservative, increasing control costs. In the context of EFC, such conservatism could lead to significant economic losses (e.g., thousands of MW in load shedding). Therefore, a control strategy that balances safety and efficiency is needed. Distributionally Robust Control (DRC) is employed to achieve this balance. DRC seeks decision outcomes under the worst-case PD of uncertain parameters, ensuring robustness while mitigating conservatism through probabilistic distribution (PD). Ref. \cite{Rahimian_2022} offers a comprehensive review of state-of-the-art DRC, including constraint conditions, the selection of ambiguity sets, and common DRO construction methods.

Currently, applying distributionally robust control to emergency frequency control is not yet widely adopted. Most existing studies remain theoretical, relying on a physical linear time-invariant (LTI) system model under model predictive control (MPC).

For linear systems, Ref. \cite{aolaritei2023wasserstein,aolaritei2023capture} constructed the empirical distribution of additive uncertainties in state-space equations using historical data. These papers define uncertainty sets based on the Wasserstein distance and analyze the analytical representation of uncertainty trajectory ranges after the temporal propagation of these sets. Ref. \cite{7122259,MARK20207136} define uncertainty sets for additive errors in linear systems using second-moment information, considering Value at Risk (VaR) and Conditional Value at Risk (CVaR), respectively, with the latter convertible to convex semidefinite programming. Ref. \cite{WU2022110648} discusses a class of distributionally robust model predictive controllers (MPC) for nonlinear stochastic processes, while Ref. \cite{9222209} proposes a distributionally robust reinforcement learning strategy, proving that the worst-case distribution is deterministic and static, thereby simplifying the problem.

For nonlinear systems, there are currently no direct distributionally robust control methods. Ref. \cite{ZHONG2023108112} linearizes the system at the operating point and then applies distributionally robust control methods designed for linear systems.

To achieve the desired properties of DRC, Gaussian Mixture Models (GMM), known for their ability to fit complex characteristics of arbitrary PDs, have been widely used to simulate uncertainties in power systems. Several studies have integrated GMM with distributionally robust optimization (DRO) in power system dispatch problems. For instance, Ref. \cite{YOU2023109087} proposed a distributionally robust optimal power flow (OPF) model for transmission grids with wind power generation, where GMM is integrated into CVaR constraints and a data-driven GMM-based ambiguity set to ensure distributional robustness. Ref. \cite{8973995} developed a DRO model for day-ahead scheduling of distribution systems with electric vehicles (EVs), using GMM to model the uncertainties of EV parameters. Additionally, Kullback-Leibler (KL) divergence is used to design the ambiguity set.

These examples show that GMM-based DRO has been extensively studied for long-term power system issues. However, applying it directly to EFC presents challenges. For instance, in Ref. \cite{YOU2023109087}, while a scalable cutting plane method is designed to improve DRO computational efficiency, it may still struggle to meet the time requirements of EFC, typically within 500ms. Furthermore, although the KL-divergence technique transforms the original DRO problem into a more computationally tractable one, DRC involves predicting future system behavior by propagating the PD of uncertain parameters through system dynamics. However, the desired propagation properties may not hold when using KL divergence to measure discrepancies between PDs. Therefore, finding a method that ensures both computational efficiency and effective propagation in DREFC remains a challenge.

In summary, current DRC methods face common challenges, such as limitations in application (e.g., requiring known linear models, affine or linear control, or light-tailed distributions). Crucially, EFC demands high control timeliness, and ensuring the computational efficiency of DRC remains an open challenge. This paper aims to address and partially resolve these issues.

\subsection{Contribution}
According to the literature review, to realize DREFC, the contributions of this paper are as follows:
\begin{enumerate}
    \item A Distributionally Robust Emergency Frequency Control framework (DREFC) is proposed to distributionally robustify the control strategies against prediction errors. This framework involves constructing the distribution of system frequency prediction errors using GMM, adopting a data-driven GMM-based ambiguity set and solving min-max optimization problem with VaR-constraints.
    
    \item We show explicitly how the Min-Max problem can be simplified into tractable minimization problems, enabling real-time implementation of the proposed DREFC framework.
    %We provide an analytical approximation for VaR constraints and reduce the computational burden of the min-max problem. For GMM-based ambiguity structures,
    
    \item The proposed DREFC can be applied to both one-shot load shedding and close-loop DC power reference regulation. For one-shot load shedding, the above framework can be directly applied. For closed-loop DC power reference regulation, we provide a method to recursively update the GMM-based ambiguity set online during the control period. This approach further improves the probability of the ambiguity set containing all distributions of prediction errors, thus ensuring frequency safety.
    % it further utilizes prediction errors during the previous timeslot to update the GMM-based ambiguity set of prediction errors for the future timeslot, reducing the conservatism of control measures.
\end{enumerate}

The paper is organized as follows: Section II-A presents a general framework for DREFC incorporating a VaR constraint. Section II-B adopts GMM to discribe the uncertainty and a Wasserstein-type distance to define the ambiguity set, which can be reformulated into a convex and explicit structure. Section II-C derives an explicit and approximate expression for the inverse cumulative distribution function (ICDF) of the GMM. Leveraging the developments in Sections II-B and II-C, Section II-D offers a computationally tractable reformulation of DREFC. In Section III, the reformulated DREFC is applied to load shedding and ancillary DC power reference regulation, with a particular focus on an online update method for the ambiguity set in the latter. Section IV validates the proposed DREFC through a case study on the CIGRE-LF system. Finally, Section V concludes the paper.

\section{Distributionally Robust EFC Formulation}
In this section, we first propose an overall framework for DREFC. Then, the formulation of the ambiguity sets and VaR constraints are introduced. Finally, a tractable reformulation of DREFC is given in the end of this section. 

\subsection{DREFC Framework}
In this subsection, the DREFC framework with VaR constraints is introduced. To explain concepts in DREFC, we will start from a classical deterministic optimization problem. Usually, an EFC based on Koopman operator described system model can be given as
\begin{subequations}
\begin{align}
  &\underset{\boldsymbol{u}}{\mathop{\min }}\,\text{  }{\boldsymbol{u}}^{\top}\boldsymbol{R}\boldsymbol{u} \\ 
  &\text{s.t.}\text{  }{\bar{f}}_{t}=\boldsymbol{C}{\boldsymbol{A}^{t}}{\boldsymbol{g}_{0}}+\boldsymbol{C}\sum\limits_{k=0}^{t-1}{\boldsymbol{A}^{t-1-k}}\boldsymbol{B}{\boldsymbol{u}_{k}},\text{  }\forall t \label{linear_dynamics}\\ 
  &\text{  }\text{  }\text{  }\text{  }{{\bar{f}}}_{t}-{{f}_{\min }}\ge \zeta ,\text{  }\forall t \label{eq_1c}\\
  &\text{  }\text{  }\text{  }\text{  }\boldsymbol{g}_{0}=\left[ \begin{matrix}
   \omega_{0}  \\
   \boldsymbol{\varphi} ({\omega }_{-\tau :0},\boldsymbol{y}_{-\tau :0})  \\
\end{matrix} \right]
\end{align}
\label{DREFC_framework}
\end{subequations}
where $\boldsymbol{g}$ is a set of finite Koopman observables which forms a subspace of the infinite dimensional Koopman observables. The subscript 0 indicates the initiation time for control. $\omega_0$ is the deviation of frequency from its nominal value (in per unit) at the initiation time for EFC, ${{\omega }_{-\tau :0}}=[{{\omega }_{-\tau }},{{\omega }_{-\tau +\Delta t}},...,{{\omega }_{0}}],{\boldsymbol{y}_{-\tau :0}}=[{{y}_{-\tau }},{{y}_{-\tau +\Delta t}},...,{{y}_{0}}]$ represent the time series of system frequency (state variable) and voltages (algebraic variables) before EFC initiates, respectively. $\boldsymbol{A}, \boldsymbol{B}, \boldsymbol{C}$ are the Koopman observables transition matrix, control matrix and output matrix. $\boldsymbol{u}$ represents the time series of control measures, expressed as $\boldsymbol{u}^{\top}=[\boldsymbol{u}_0^{\top}, \boldsymbol{u}_1^{\top},...,\boldsymbol{u}_{t-1}^{\top}]$.
With the loss function and the algorithm provided in Ref. \cite{cao2023datadrivenfrequencyloadshedding}, it is feasible to approximate the parameters of $\boldsymbol{g}$, if the basis function for $\boldsymbol{g}$ is predefined, along with the matrices $\boldsymbol{A}$ and $\boldsymbol{B}$. $\boldsymbol{C}$ is a predefined vector. $\bar{f}_{t}$ denotes system frequency at time $t$ predicted by Koopman linear system, ${f}_{\text{min}}$ is the minimal allowed system frequency. $\zeta$ is a safety margin. For more details on Eq. \eqref{linear_dynamics}, see the author's previous research \cite{cao2023datadrivenfrequencyloadshedding,10135149}.

In Eq. \eqref{DREFC_framework}, the system frequency is predicted by Eq. \eqref{linear_dynamics}. However, the predicted frequency may have a small error compared to the actual future frequency. This is due to noisy input in $\boldsymbol{g}_0$ and training error in $\boldsymbol{g}$, $\boldsymbol{A}$, $\boldsymbol{B}$. Therefore, an uncertain discrepancy between predicted and real frequency is unavoided. Hence, $\zeta$ in Eq. \eqref{eq_1c} is introduced to avoid that the real future frequency fall below the allowed minimum value for under-frequency event. One way is to determine the margin by the largest prediction error of system frequency in the set of history prediction errors, which makes the design of the safety margin a RO problem. However, for EFC, RO results in relatively conservative control measures. To reduce the conservatism, the discrepancy can be regarded as a disturbance to the nominal model Eq. \eqref{linear_dynamics}, and the disturbance is governed by an unknown distribution. This has led to the need for method which are robust with respect to the unknown PD of the disturbance.

In most applications, PDs are not directly observable and predictable, and must be estimated from data. Specifically, suppose one has access to samples from a real underlying distribution $\hat{\mu}$. Since only partial statistical information about the distribution is available, we can only construct a reference distribution $\mu$ from the available data. DRO methods typically assume that the distribution $\hat{\mu}$ lies in some set of distributions ${\Omega }_{\mu}$, which is centered at a reference distribution $\mu$. The optimization is then done w.r.t. the worst case distribution in this set. Then, the optimization problem can be reformulated into then a min-max distributionally robust optimal EFC problem as follows, if the VaR constraint is considered

\begin{subequations}
\begin{align}
  \textbf{DREFC-F:} \quad& \underset{\boldsymbol{u}}{\mathop{\min }}\,\text{  }{{\boldsymbol{u}}^{\top}}\boldsymbol{R}\boldsymbol{u} \\
 & \text{s}\text{.t}\text{.   }{{{\bar{f}}}_{t}}=\boldsymbol{C}\boldsymbol{A}^{t}\boldsymbol{g}_{0}+\boldsymbol{C}\sum\limits_{k=0}^{t-1}\boldsymbol{A}^{t-1-k}\boldsymbol{B}{\boldsymbol{u}_{k}},\text{  }\forall t \label{eq_2b}\\ 
 & \text{  }\text{  }\text{  } {{\bar{f}}_{t}}-{{f}_{\min }}\ge \underset{{\hat{\mu }}}{\mathop{\text{max}}}\,\text{ICDF}_{{\hat{\mu }}}(1-\alpha ),\text{  }\forall t \label{eq_2c}\\ 
 & \text{  }\text{  }\text{  } {\hat{\mu }}\in {{\Omega }_{{{\mu }}}} \label{eq_2d}
\end{align}
\label{DREFC_framework_2}
\end{subequations}
where $\mu$ represents a reference PD estimated from data, $\hat{\mu}$ is an unknown distribution that belongs to the ambiguity set ${\Omega }_{{\mu }}$. ICDF is the inverse cumulative distribution function, $\alpha$ represents the significance level. Eq. \eqref{eq_2c} finds the worst $\hat{\mu}$ in the ambiguity set ${\Omega }_{{\mu }}$, maximizing the safety margin calculated by $\text{ICDF}(1-\alpha)$. When the PD of system frequency prediction errors is $\hat{\mu}$, the VaR contraint Eq. \eqref{eq_2c} ensures the system frequency safety probability is $1-\alpha$.

DREFC-F provides a general formulation for DREFC, in which Eq. \eqref{eq_2c} and Eq. \eqref{eq_2d}  are implicit and need further elaboration to ensure the computational feasibility of DREFC-F. The challenges we face are twofold:
\begin{enumerate}
 \item How to describe $\mu$ to accurately represent the complex characteristics of frequency prediction errors while easing subsequent description of ambiguity set and ICDF.
 % \item How to describe ${\Omega }_{{\mu}_{0}}$ to ensure the overall optimization problem remains tractable.
 \item VaR-based chance constraints Eq. \eqref{eq_2c} are often non-linear and non-convex, making them computationally intractable. Making these constraints computationally tractable is challenging.
\end{enumerate}

This paper will address the above challenges in the following sections.

\vspace{0.2cm}
\noindent \textbf{Remark.} Overall, both VaR and CVaR have their rationality and the decision maker may prefer one over the other in different situations. In the DREFC framework, the use of VaR is intuitive since it limits the violation probability of the frequency safety constraint.

% \vspace{0.2cm}
% \noindent \textbf{Remark.} According to Ref. \cite{}, prediction errors的成因通常较为复杂，包括但不限于量测误差、数据bias、训练误差等等，本文由于主要对$\zeta$进行设计，$\zeta$的设计旨在使得实际系统频率高于$f_min$，因此其主要用于衡量实际系统频率和利用Eq.\eqref{}得到的频率的差值即可。且本文假设training set和testing set具备相同数据在利用training set 的trajectory因此本文没有详细区分什么原因导致的频率预测误差，而是统一使用仅使用加性误差

\subsection{Describing $\mu$ and ${\Omega }_{{\mu}_{0}}$}
\label{Describing_mu_0_and_Omega_mu0}
In this subsection, a data-driven GMM-based ambiguity set is used to hedge against the uncertainty in the distribution of frequency prediction error. First, we use GMM to build the joint PD of frequency prediction error according to its historical data. That is, the PDF of the prediction error is defined as the following GMM

\begin{align}
    \text{PDF}(\tilde{\xi}|\Psi) = \sum_{k=1}^{K} \pi_k \phi(m_k, \Sigma_k)
\end{align}
where $\phi(m_k, \Sigma_k)$ is the multivariate Gaussian distribution with the mean $m_k$ and covariance matrix $\Sigma_k$, and it is the $k$th Gaussian component of the GMM; $(\hat{\pi}, \dots, \pi_K)$ are the mixing weights of the $K$ Gaussian components; $\Psi$ is the vector form of $\{\hat{\pi}, \dots, \pi_K, m_1, \dots, m_K, \Sigma_1, \dots, \Sigma_K\}$.

Given the historical data of $\tilde{\xi}$, the parameters $K$ and $\Psi$ can be obtained by the maximum likelihood estimation (MLE) \cite{5298967}.

The reasons that we adopt GMM are fourfold.

\begin{enumerate}
    \item The PD of prediction errors is non-Gaussian, exhibiting skewness, multi-peaks, and heavy tails. GMM can fit these complex characteristics of PDs \cite{LI20161530}.

    \item Though the GMM-based VaR lacks an analytical form, the use of GMM makes it possible to approximate the ICDF, which makes VaR analytical \cite{7862254}.
    
    \item GMM can capture the correlation between random variables \cite{WANG2018771}, and its marginal distributions are also GMMs. These mathematical features of GMM facilitate the online updating of $\mu$.
    
    \item A Wasserstein-type distance on GMM can be introduced, providing a convex and explicit formulation that depends only on the parameters of the Gaussian distributions in the mixture.
\end{enumerate}

According to Ref. \cite{doi:10.1137/19M1301047}, let $\mu = \sum_{k=1}^{K} \pi^k \mu^k$ and $\hat{\mu} = \sum_{l=1}^{L} \hat{\pi}^l \hat{\mu}^l$ be two GMM, then
\begin{align}
    MW_2^2(\mu, \hat{\mu}) = \min_{w \in \Pi(\pi, \hat{\pi})} \sum_{k,l} w_{kl} W_2^2(\mu^k, \hat{\mu}^l) \label{wtd_two_GMM}
\end{align}

where 
\[
W_2^2(\mu, \hat{\mu}) = \|m - \hat{m}\|^2 + \text{tr} \left( \Sigma + \hat{\Sigma} - 2 \left( \Sigma^{\frac{1}{2}} \hat{\Sigma} \Sigma^{\frac{1}{2}} \right)^{\frac{1}{2}} \right)
\]
and 
\begin{align}
    \Pi(\pi, \hat{\pi}) = \{ w \in &\mathcal{M}_{K, L}(\mathbb{R}^+); \, \forall k,\notag \\
&\sum_l w_{k,l} = \pi^k \text{  }, \text{and} \text{  } \forall l, \sum_k w_{k,l} = \hat{\pi}^l \}
\end{align}

If we use the above Wasserstein-type distance to determine the ambiguity set, the set of $MW_2^2(\mu, \hat{\mu})\le \gamma$ can be reformulated as follows
\begin{subequations}
\begin{align}
\text{  }\text{  } \sum_{k,l} \omega_{k,l} &\left[(m^l - \hat{m}^k)^2 + (\sigma^l - \hat{\sigma}^k)^2 \right] \leq \gamma \\
\sum_k \omega_{k,l} &= \pi^l \\
\sum_l \omega_{k,l} &= \hat{\pi}^k 
\end{align}
\label{eq_ambiguity}
\end{subequations}
where $\gamma$ is a preset value for the radius of ambiguity set and
\begin{subequations}
\begin{align}
    \mu =  \sum_{l=1}^{L} \pi^l \phi(m^l, (\sigma^l)^2)\label{eq_4a}\\
\hat{\mu} = \sum_{k=1}^{K} \hat{\pi}^k \phi(\hat{m}^k, (\hat{\sigma}^k)^2)\label{eq_4b}
\end{align}
\label{eq_gmm}
\end{subequations}
%\mu^l, \quad \mu^l \sim N(m^l, (\sigma^l)^2) 

The Wasserstein distance can be seen as the minimum transport cost of the probability mass of one distribution into another. Eq. \eqref{eq_ambiguity} provides a convex expression for the Wasserstein-type distance between underlying PD and the reference distribution. In next subsection, we'll discuss how to find the worst-case risk over the ambiguity set.

%性质
\subsection{Approximation of VaR contraint}
\label{Approximation_of_VaR_contraint}

After the elaboration of the ambiguity set, we propose an approximation method for ICDF to make the VaR constraint computationally efficient. 

For the VaR constraint Eq.\eqref{eq_2c} in DREFC, the non-analytic nature arises primarily from the non-analyticity of the ICDF. Ref. \cite{7862254} suggests a method to approximate the cumulative distribution function of GMM using a fourth-order polynomial to make it analytical. However, further deriving the ICDF involves solving the polynomial, which is highly complex. Ref. \cite{YOU2023109087} proposed an iterative cutting plane-based solution algorithm to enforce the CVaR constraints in optimal power flow (OPF) problems, yet its computational efficiency may not meet the real-time demands of EFC, typically requiring millisecond-level response times. Consequently, this subsection introduces an approximation method for the ICDF, allowing it to be expressed in terms of the parameters of GMM.

The approximated ICDF of GMM is:
\begin{align}
    \text{icdf}_{\hat{\mu}}(1-\alpha) \approx \sum_{k=1}^{K} \hat{\pi}^k \left( \hat{m}^k + \hat{\sigma}^k \cdot Z_{1-\alpha} \right) \label{eq_approximated_icdf}
\end{align}

Here, $\hat{m}^k + \hat{\sigma}^k \cdot Z_{1-\alpha}$ is an accurate expression for PD of the $k$-th Gaussian component of GMM $\hat{\mu}^{k}$. Eq.\eqref{eq_approximated_icdf} constructs the ICDF of the GMM by weighting the ICDF of each Gaussian component according to their weights. Since this is an approximation for the GMM's ICDF, in Section \ref{Test_Approximation}, we will experimentally verify the potential errors of this approximation and test how these errors affect the control results of DREFC, such as whether they reduce the number of frequency-safe scenarios.

\subsection{Reformulation of DREFC}

Based on Section \ref{Describing_mu_0_and_Omega_mu0} and Section \ref{Approximation_of_VaR_contraint}, the original implicit and non-convex problem in Eq.\eqref{DREFC_framework} becomes an explicit and computationally efficient problem, as given in DREFC-R.

\begin{align}
&\textbf{DREFC-R:} \notag\\
& \min_{\boldsymbol{u}} \; \boldsymbol{u}^{\top} \boldsymbol{R} \boldsymbol{u} \notag\\
& \text{s.t.} \text{    Eq. \eqref{eq_2b}, Eq. \eqref{eq_ambiguity},\text{  }Eq. \eqref{eq_gmm}} \notag\\
% \quad \overline{f}_t = \boldsymbol{C}\boldsymbol{A}^t \boldsymbol{g}_0 + \boldsymbol{C} \sum_{k=0}^{t-1} \boldsymbol{A}^{t-1-k} \boldsymbol{B} \boldsymbol{u}_k, \forall t=1,2,...,T\\
& \overline{f}_t - f_{\min} \geq \max_{\hat{\mu}} \sum_{k=1}^{K} \hat{\pi}^k \left( \hat{m}^k + \hat{\sigma}^k \cdot Z_{1-\alpha} \right), \forall t=1,2,...,T  
% &\sum_{k,l} \omega_{k,l} \left[(m^l - \hat{m}^k)^2 + (\sigma^l - \hat{\sigma}^k)^2 \right] \leq \gamma \\
% &\sum_k \omega_{k,l} = \pi^l \\
% &\sum_l \omega_{k,l} = \hat{\pi}^k 
\end{align}

In the problem above, since decision variables $\boldsymbol{u}_k$ and $\bar{f}_t$ in Eq. \eqref{eq_2b} and Eq. \eqref{eq_2c} do not affect the worst-case distribution in the ambiguity set described by Eq. \eqref{eq_ambiguity}, DREFC-R can be divided into two subproblems: an lower problem and an upper problem, which can be expressed as:
\begin{flalign*}
\textbf{DREFC-L:} \text{   }\text{   }\text{   }&\text{   }\text{   }\text{   }\text{   }\text{   }\text{   }\text{   }\text{   }\text{   }\text{   }\text{   }\text{   }\text{   }\text{   }\\
\quad  &\max_{\hat{\mu},\zeta} \sum_{k=1}^{K} \hat{\pi}^k \left( \hat{m}^k + \hat{\sigma}^k \cdot Z_{1-\alpha} \right) \text{   }\text{   }\text{   }\text{   }\text{   }\text{   }\text{   }\text{   }\text{   }\text{   }\text{   }\\
%
% & \text{Eq.\eqref{eq_ambiguity},\text{  }Eq. \eqref{eq_gmm}}
% \text{s.t.}\text{  }\text{  } \sum_{k,l} \omega_{k,l} &\left[(m^l - \hat{m}^k)^2 + (\sigma^l - \hat{\sigma}^k)^2 \right] \leq \gamma \\
% \sum_k \omega_{k,l} &= \pi^l \\
% \sum_l \omega_{k,l} &= \hat{\pi}^k \\
% \hat{\mu} =  \sum_{k=1}^{K}& \hat{\pi}^k \phi(\hat{m}^k, (\hat{\sigma}^k)^2)
&\text{s.t.}\text{   }\text{   }\text{Eq. \eqref{eq_ambiguity},\text{  }Eq. \eqref{eq_gmm}}  \notag
\end{flalign*}

\begin{flalign*}
\textbf{DREFC-U:}\text{   }\text{   }\text{   }\text{   }&\text{   }\text{   }\text{   }\text{   }\text{   }\text{   }\text{   }\text{   }\text{   }\text{   }\text{   }\text{   }\text{   }\text{   }\text{   }\text{   }\text{   }\text{   }\text{   }\text{   }\text{   }\text{   }\text{   }\text{   }\text{   }\text{   }\text{   }\text{   }\text{   }\text{   }\text{   }\text{   }\text{   }\text{   }\text{   }\text{   }\text{   }\text{   }\text{   }\text{   }\text{   }\text{   }\text{   }\text{   }\text{   }\text{   }\text{   }\text{   }\text{   }\text{   }\text{   }\text{   }\text{   }\text{   }\text{   }\text{   }\text{   }\text{   }\text{   }\text{   }\text{   }\text{   }\text{   }\\
&\min_{u} \quad  \boldsymbol{u}^{\top} \boldsymbol{R} \boldsymbol{u} \\
&\text{s.t.} \quad  \text{Eq. \eqref{eq_2b}}\\
& \overline{f}_t - f_{\min} \geq \zeta^{*}, \forall t=1,2,...,T 
\end{flalign*}
where $\zeta^{*}=\arg\max_{\zeta} \sum_{k=1}^{K} \hat{\pi}^k \left( \hat{m}^k + \hat{\sigma}^k \cdot Z_{1-\alpha} \right)$ in DREFC-U.

With DREFC-L and DREFC-U, DREFC collapses to computationally tractable finite-dimensional optimization problems. Specifically, DREFC-U can be solved in polynomial time due to the positive definiteness of $\boldsymbol{R}$ in the objective function. In contrast, DREFC-L has a bilinear objective function, causing its computational complexity to increase exponentially with the values of $K$ and $L$. However, since $K$ and $L$ represent the number of Gaussian mixture components in the GMM, they are generally not large—typically less than 10 or fewer in practical applications. Therefore, the computation time remains manageable. 

Unlike the method proposed in Ref. \cite{aolaritei2023capture} and other standard approaches, where the computational load increases with the amount of historical data, this method theoretically maintains consistent computational efficiency because the growth in data does not directly affect the number of parameters in the GMM. Since the computational complexity of DREFC is mainly influenced by the number of parameters in the GMM, the amount of historical data does not impact the computational efficiency of DREFC.

% \vspace{0.2cm}
% \noindent \textbf{Remark.} (Computational Complexity) 

\section{DREFC for different control resources}
In Section II, we designed a framework for emergency frequency control that limits the violation probability of the frequency safety margin under the true distribution of frequency prediction errors, while ensuring computational efficiency. When applying DREFC to different control resources, it can be tailored to the specific characteristics of each resource. In this section, we will discuss these applications in detail.
\subsection{Load shedding}
For EFC, load shedding can be designed as a one-shot action since step-by-step protection actions lead to time delays \cite{Xu2015}. Therefore, the distributionally robust load shedding amount can be calculated at the beginning of the under-frequency event. Specifically, the reference distribution of frequency nadir $\mu$ is calculated with offline samples of frequency nadir prediction errors obtained at the end of training for linear dynamics Eq. \eqref{linear_dynamics}. Then, DREFC-L and DREFC-U is calculated sequentially. 

In this case, since the reference distribution only has to be calculated for only once offline, DREFC-L can be calculated offline to obtain $\zeta^{*}$. When the frequency drops below a certain threshold, DREFC-U is computed online. Since DREFC-U polynomial time complexity, it ensures efficient real-time computation.

\subsection{Ancillary DC power reference regulation}
\label{Ancillary_DC_power_reference_regulation}

Adjusting the DC power reference offers a fast, adaptable resource, allowing continuous control adjustments based on system measurements until frequency recovery \cite{10049715}. Unlike load shedding, for which frequent adjustments are not practical, DC power reference can be continuously adjusted in a moving horizon fashion. Therefore, adjusting the DC power reference is often designed as a closed-loop control strategy\cite{cao2023data,10049715}. Hence, DREFC with DC power as the control resource can also be designed as a closed-loop control strategy. For moving time window, for each time window, the data of the past time windows can be collected to provide more information for a more precise description of data-driven GMM-based ambiguity set.

The mathematical properties of GMM to capture the correlation between random variables help utilize past time window data (collected online) to update the reference distribution of prediction errors in the near future, thereby updating the ambiguity set. This approach improves the precision of the ambiguity set construction in real-time, enhancing the probability of efficiently capturing distributional uncertainty within an ambiguity set. The method for online updates is as follows:

\vspace{0.2cm}
\noindent \textbf{Phase 1:} Let a random vector $X_p=[x_{t-M},x_{t-M+1},…,x_{t}]^{\top}$ denote a time series of frequency errors of $M$ steps before time $t$, $X_f = [x_{t+1},x_{t+2},…, x_{t+N}]^{\top}$ denote frequency errors of $N$ step after time $t$. A GMM is used to represent the joint PDF of an aggregated random vector $[X_p^{\top}\ X_f^{\top}]^{\top}$.

\vspace{0.2cm}
\noindent \textbf{Phase 2:} Constructing conditional distributions of future prediction errors with respect to past prediction errors $[X_f\mid X_p]$. $[X_f\mid X_p]$ represents represents the conditional distribution of the future prediction errors $X_f$ given past prediction errors $X_p$.

\vspace{0.2cm}
\noindent \textbf{Phase 3:} When DREFC for DC power reference regulation is implemented online, a trajectory of frequency errors $X_p$ can be calculated for past $M$ steps. Then the GMM parameter set of $[X_f\mid X_p]$ can be updated as follows \cite{WANG2018771}.

\begin{align}
f_{X_f|X_p}(x_f|x_p) &= \sum_{g=1}^{G} \omega'_g N_g(x_f|x_p; \mu^{x_f \cdot x_p}_g, \sigma^{x_{f}x_f \cdot x_p}_g)
\end{align}
where
\begin{align}
    &\omega'_{g} = \omega_g \frac{N_g(x_{p}; \mu^{x_{p}}_g, \sigma^{x_{p}x_{p}}_g)}{\sum_{l=1}^{G} \omega_l N_l(x_{p}; \mu^y_l, \sigma^{x_{p}x_{p}}_l)} \\
&\mu^{x_f \cdot x_{p}}_g = \mu^{x_{f}}_g + \sigma^{x_{f}x_{p}}_g (\sigma^{x_{p}x_{p}}_g)^{-1} (x_{p} - \mu^{x_{p}}_g) \\
&\sigma^{x_{f}x_f \cdot x_{p}}_g = \sigma^{x_{f}x_{f}}_g - \sigma^{x_{f}x_{p}}_g (\sigma^{x_{p}x_{p}}_g)^{-1} \sigma^{x_{p}x_{f}}_g 
\end{align}
where $f_{X_{f}X_{p}}(x_{f},x_{p})$ is the joint PDF of $[X_{f}^{\top} \ X_{p}^{\top}]^{\top}$; $N_g(\cdot), N_l(\cdot)$ denotes a multivariate Gaussian distribution function, which is called the $g$th and $l$th Gaussian component of the GMM; $G$ is the total number of Gaussian components; $\mu$ and $\sigma$ with subscripts and superscripts are parameters.

In the implementation procedure, the GMM parameter set of $[X_p^{\top}\ X_f^{\top}]^{\top}$ is estimated according to historical data in the first phase. That is, the parameter set of a GMM is obtained offline. During the online process, $\mu$ can be updated in a moving horizon. Specifically, At time $t$, $X_p$ can be obtained by calculating the deviation of the frequency calculated with Koopman linear prediction error and measured frequency. Then the GMM parameter set of $[X_f\mid X_p]$ can be updated. The updated PDF of $[X_f\mid X_p]$ is served as the input for DREFC-L.

\section{Case Study}
In this section, we validate the effectiveness of the proposed DREFC. First, we evaluate GMM's accuracy in modeling prediction errors over a specific period and the conditional PD of these errors. Next, we analyze the approximation error of the ICDF introduced in Section \ref{Approximation_of_VaR_contraint} and its potential impact on the distributionally robust control strategy. Finally, we test DREFC's performance in terms of safety, efficiency, and timeliness.

\subsection{Test System Description}
To validate the effectiveness of our proposed method, we conducted simulation experiments on the CEPRI-FS test system. The electromechanical transient model for this case study is available for access at Ref. \cite{CSEE-Benchmark2024}. Similar to Ref. \cite{10135149}, which also uses Koopman operator described system model with data, this study generates 300 frequency trajectories of 60s as the training set. Therefore, we could obtain 300 trajectories of prediction errors, which can be used for obtaining the reference GMM parameter set.
This section will conduct evaluations of the modeling accuracy of the GMM and the approximation accuracy of ICDF in Eq. \eqref{eq_approximated_icdf}.

\subsection{Modelling accuracy of GMM}
\label{Modelling_accuracy_of_GMM}
\subsubsection{Frequency nadir}
\begin{figure}[t] %可选参数 h t b p，代表允许图片出现的位置，h表示此处附近，t表示顶部，b表示底部，p表示单独一页，H表示固定此处
    \centering
    % \vspace{-0.2cm}  %调整图片与上文的垂直距离
    % \setlength{\abovecaptionskip}{-0.05cm}   %调整图片标题与图距离
    % \setlength{\belowcaptionskip}{-2cm}   %调整图片标题与下文距离
    \includegraphics[width=8cm]{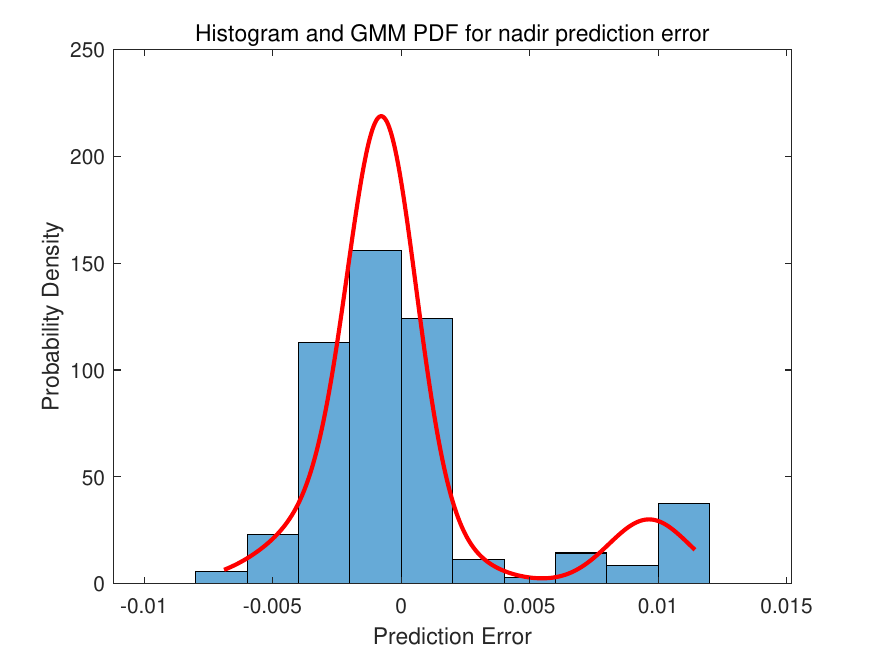}
    \caption{Histogram and GMM PDF for PEFNs} \label{conditional_plotting_1}
\end{figure}

The prediction errors of frequency nadir (PEFN) from the training datasets are used for this test. A GMM with three components is applied to model the PEFN distribution across different trajectories, and the results are compared with prediction error histograms. As shown in Fig. \ref{conditional_plotting_1} , the GMM distributions align well with the histograms. To evaluate accuracy, the root-mean-square error (RMSE) of the GMM and other non-Gaussian modeling methods, including Gaussian Copula and t Copula, are calculated. The RMSE values are 0.11, 0.25, and 0.69, respectively, indicating the GMM's superior accuracy in modeling complex distributions.

\subsubsection{Conditional distributions}
Since the ancillary DC power reference regulation method in Section \ref{Ancillary_DC_power_reference_regulation} requires online updates of the reference PD, which is designed based on the conditional PD of a GMM, we first visualize the joint probability density of frequency prediction errors at two time points to illustrate the correlation between them. Additionally, we verify the improvement in prediction accuracy for future errors given the knowledge of prior prediction errors.

The data of prediction errors for the frequency trajectories at $t = 1,2,…,T$ in the training datasets is used for this test. Then, we construct four random variables with $\boldsymbol{X}_1=[x_1,x_2,…,x_{T-1}]^{\top}$, $\boldsymbol{Y}_1=[x_2,x_3,…,x_{T}]^{\top}$ and $\boldsymbol{X}_2=[x_1,x_2,…,x_{T-2}]$ and $\boldsymbol{Y}_2=[x_3,x_4,…,x_{T}]$. The joint distribution of $[\boldsymbol{X}_1^{\top}\ \boldsymbol{Y}_1^{\top}]^{\top}$ and $[\boldsymbol{X}_2^{\top} \ \boldsymbol{Y}_2^{\top}]^{\top}$ is modeled by the GMM with 3 components.

\begin{figure*}[t] %可选参数 h t b p，代表允许图片出现的位置，h表示此处附近，t表示顶部，b表示底部，p表示单独一页，H表示固定此处
    \centering
    % \vspace{-0.2cm}  %调整图片与上文的垂直距离
    % \setlength{\abovecaptionskip}{-0.05cm}   %调整图片标题与图距离
    % \setlength{\belowcaptionskip}{-2cm}   %调整图片标题与下文距离
    \includegraphics[width=16cm]{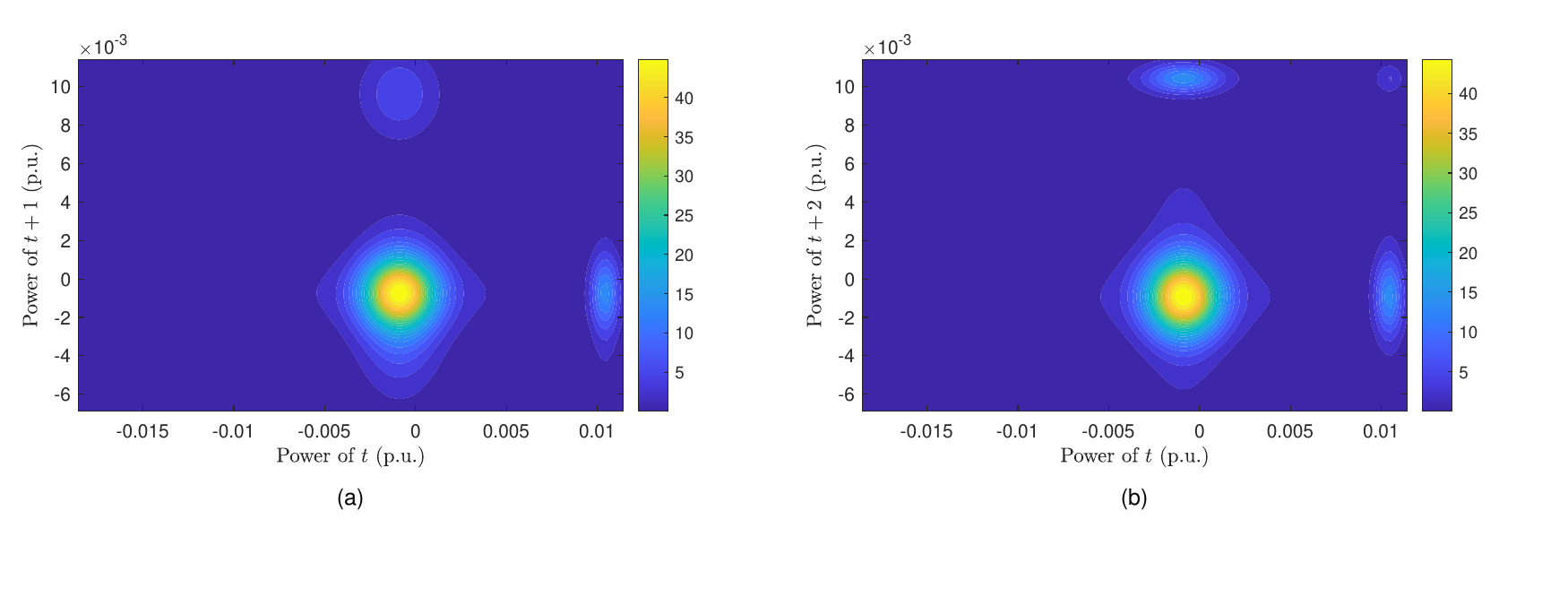}
    \caption{Conditional joint distributions of frequency prediction error at $t$, $t+1$ and $t$, $t+2$.} \label{conditional_plotting_combined}
\end{figure*}

An illustrative example of the joint distribution of prediction errors at two different time points is provided in Fig. \ref{conditional_plotting_combined}. For instance, in Fig. \ref{conditional_plotting_combined}(b), when the prediction error at time $t$ falls within the range of -0.001 to 0, the prediction error at $t+2$ is modeled by a GMM with two components. When the prediction error at time $t$ is between 0 and 0.01, the $t+2$ prediction error is dominated by a single Gaussian component. Similarly, in Fig. \ref{conditional_plotting_combined}(a), when the prediction error at time $t$ is between -0.001 and 0, the $t+2$ prediction error is represented by a GMM with at least two components, while in the range of 0 to 0.01, it is modeled by a GMM with two components but different parameters. This shows that prediction errors at different time points are interdependent. It is important to note that Fig. \ref{conditional_plotting_combined} visualizes this interdependence by plotting prediction errors at two time points as the axes, emphasizing their correlation. In practice, the method in Section \ref{Ancillary_DC_power_reference_regulation} uses data from a prior time period to update the GMM parameters for predicting future errors, not just for two specific time points.

Next, we verified the improvement in the accuracy of predicting subsequent PEFN when prior prediction errors are known. To demonstrate this, we set $M=1, N=1$ for $X_p$ and $X_f$. We utilized 300 trajectories of PEFN from the training set to construct multiple sets of time-series data $[x_{t-1}, x_{t}, x_{t+1}]$. Using these time-series data, we generated the probability density functions $PD(x_{t+1})$ and $PD(x_{t+1}|x_{t-1}, x_{t})$ with GMM. For a specific set of time-series values $[x_{t-1}, x_{t}, x_{t+1}]$, we calculated the probability of $x_{t+1}$ occurring using both PDs. It was found that the probability of $x_{t+1}$ calculated using $PD(x_{t+1}|x_{t-1}, x_{t})$ was, on average, approximately 0.8 higher than when calculated using $PD(x_{t+1})$, indicating greater accuracy in capturing the subsequent prediction error.

\subsection{Approximation of ICDF}
\label{Test_Approximation}
To evaluate the approximation accuracy of the ICDF in Eq. \eqref{eq_approximated_icdf}, we generate 100 PDs of GMM with random parameters. At a confidence level of 0.95, we calculated both the true ICDF values of these GMMs and the approximated values from Eq. \eqref{eq_approximated_icdf}. The two plots show a strong linear correlation between the true ICDF values and the approximations, with Pearson correlation coefficients of 0.98 and 0.96, respectively. To further support the use of max $\sum_{k=1}^{K} \hat{\pi}^k \left( \hat{m}^k + \hat{\sigma}^k \cdot Z_{1-\alpha} \right)$ as a substitute for max $\text{icdf}_{\hat{\mu}}(1-\alpha)$, we also calculated the probability that the approximation maintains the comparison between the ICDFs of any two GMMs. The results, shown in Table \ref{Table_of_Results}, indicate a probability of over 99.5\%.

As shown in Table \ref{Table_of_Results}, in at least 99.5\% of cases, max $\sum_{k=1}^{K} \hat{\pi}^k \left( \hat{m}^k + \hat{\sigma}^k \cdot Z_{1-\alpha} \right)$ can replace max $\text{icdf}_{\hat{\mu}}(1-\alpha)$. For the remaining 0.5\% cases, the proposed method converges to a suboptimal solution, which may result in a slightly lower frequency safety probability than the target confidence level of $1-\alpha$ when the frequency prediction error follows the worst-case distribution. However, as demonstrated by the subsequent safety tests, the frequency safety probability remains close to the set confidence level, even in worst-case scenarios.
\begin{table}[htbp]
    \centering
    \caption{The probability that the approximation maintains the comparison between the ICDFs of any two GMMs}
    \small
    \begin{tabular}{cccccc}
        \toprule
        $K$ & $\alpha=0.01$ & $\alpha=0.05$ & $\alpha=0.1$ & $\alpha=0.15$ & $\alpha=0.2$ \\
        \midrule
        3 & 0.38\% & 0.20\% & 0.10\% & 0.08\% & 0.13\% \\
        4 & 0.40\% & 0.20\% & 0.10\% & 0.08\% & 0.14\% \\
        5 & 0.00\% & 0.23\% & 0.13\% & 0.08\% & 0.16\% \\
        \bottomrule
    \end{tabular}
    \label{Table_of_Results}
\end{table}
\subsection{Frequency Safety}
This section first tests whether DREFC can ensure frequency safety for load shedding and ancililary DC power regulation.
\subsubsection{Load Shedding}
Theoretically, DREFC is designed to ensure frequency safety with a certain confidence level, even under the worst-case distribution of frequency errors. To validate its control effectiveness, we used PEFN from the training dataset to learn the parameters of the reference GMM. With learnt parameters of the reference GMM, 700 possible prediction errors are generated according to methods in Ref. \cite{WANG2018771}. Then the 700 prediction errors are applied to the linear prediction system in Eq. \eqref{linear_dynamics}, forming 700 scenarios. Random generator failures were triggered to create low-frequency events in these scenarios, and DREFC was used to determine the load shedding amount for each scenario. We set $\alpha$ to 0.05 and $f_{min}$ to -0.015. 

\begin{figure}[t] %可选参数 h t b p，代表允许图片出现的位置，h表示此处附近，t表示顶部，b表示底部，p表示单独一页，H表示固定此处
    \centering
    % \vspace{-0.2cm}  %调整图片与上文的垂直距离
    % \setlength{\abovecaptionskip}{-0.05cm}   %调整图片标题与图距离
    % \setlength{\belowcaptionskip}{-2cm}   %调整图片标题与下文距离
    \includegraphics[width=8cm]{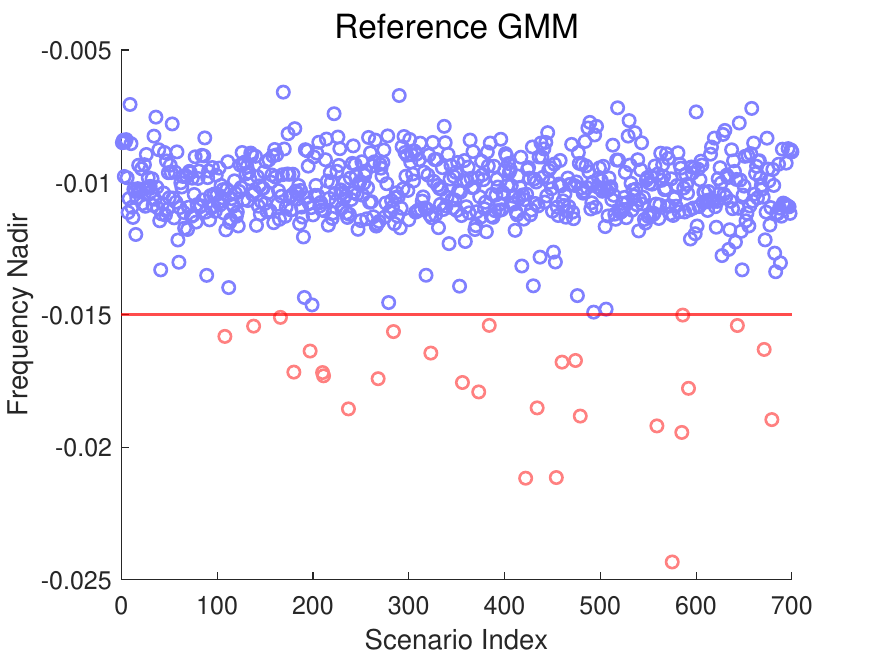}
    \caption{Frequency nadir of 700 scenarios with proposed DREFC method when the reference distribution is used to generate the possible PEFNs.} \label{DRO3_1}
\end{figure}

With reference distribution of prediction errors, DREFC was expected to keep the system frequency above -0.015 over 95\% of cases since the reference distribution is often not the worst distribution. Fig. \ref{DRO3_1} shows the frequency nadirs across 700 scenarios generated under the reference distribution, with those above $f_{min}$ marked in purple and those below in red. The results indicate that system frequency remained safe in 96\% of scenarios, exceeding the 95\% target.
\begin{figure}[t] %可选参数 h t b p，代表允许图片出现的位置，h表示此处附近，t表示顶部，b表示底部，p表示单独一页，H表示固定此处
    \centering
    % \vspace{-0.2cm}  %调整图片与上文的垂直距离
    % \setlength{\abovecaptionskip}{-0.05cm}   %调整图片标题与图距离
    % \setlength{\belowcaptionskip}{-2cm}   %调整图片标题与下文距离
    \includegraphics[width=8cm]{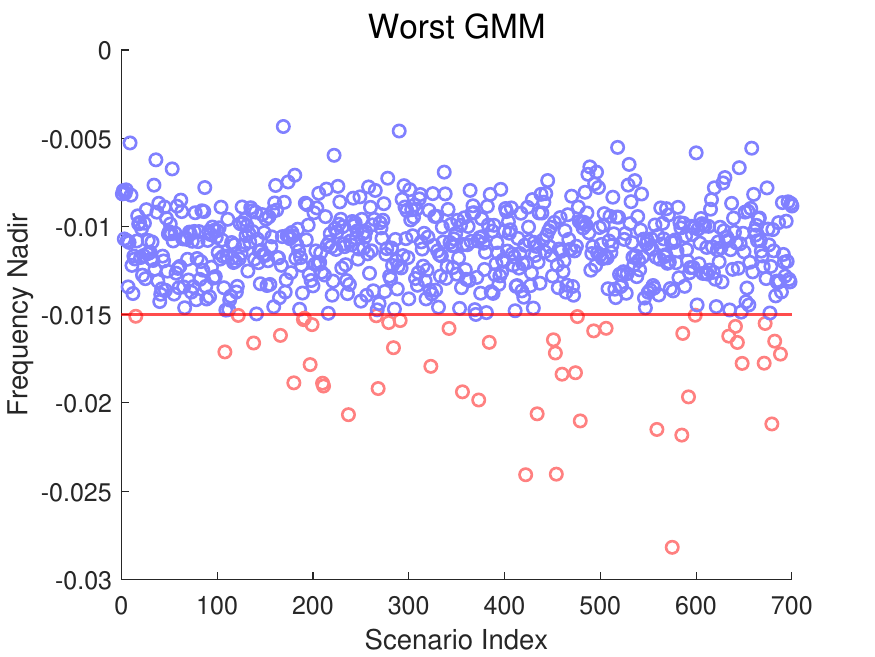}
    \caption{Frequency nadir of 700 scenarios with proposed DREFC method when the worst distribution is used to generate the possible PEFNs.} \label{DRO3_2}
\end{figure}

Under the worst distribution of prediction errors, DREFC was expected to keep the system frequency above -0.015 in 95\% of cases. However, since the approximation of ICDF leads to suboptimal solutions in a small portion of scenarios (often less than 0.5\%), DREFC may keep the system frequency above -0.015 for cases a little bit less than 95\%. To verify this point, we used the worst-case GMM calculated from DREFC-L to generate 700 possible prediction errors and tested whether the system frequency could be safe in 95\% of cases. 

Fig. \ref{DRO3_2} shows the frequency nadirs across the 700 scenarios. The results indicate that the system frequency was safe in 94.7\% of scenarios, slightly lower than the preset confidence level of 95\%. This is due to the ICDF approximation error. In this case, to ensure that under the worst distribution of prediction errors, DREFC is expected to maintain the system frequency above -0.015 for a preset probability level, we can set $1-\alpha$ in DREFC-L slightly higher than the preset probability level. For instance, if we want DREFC to keep the system frequency above -0.015 in 95\% of scenarios, we can set $1-\alpha$ slightly above 95\%. For example, the ICDF approximation error affects the optimal solution with a probability of less than 0.5\% (as discussed in Section IV.C), we can therefore set $1-\alpha$ to 95.5\%.

\subsubsection{DC power regulation}
Compared to one-time load shedding, DC power regulation is more suitable for rapid adjustment throughout the entire duration of under frequency events. Therefore Section \ref{Ancillary_DC_power_reference_regulation} introduces a method for the online update of PEFN to enhance the accuracy of the reference distribution.

This section employs both safety and economic indicators to compare the advantages of the online update DREFC with the non-online DREFC. The security indicator remains consistent with the previous section, where the probability of the system frequency staying within the safe range across 700 scenarios is used. 

Furthermore, we define an economic indicator based on the probability that the system’s nadir frequency exceeds a preset value across 700 scenarios. The reason is given as follows. Usually, increasing the amount of load shedding typically leads to a higher system frequency. However, if the frequency remains within the safe range, a greater deviation of the nadir and steady-state value of the system frequency from the specified hard limits incurs higher control costs. Considering that the safety threshold for frequency in this paper is set at greater than -0.015, a minimum economic frequency threshold of -0.005 is selected. This threshold means that when the frequency nadir exceeds -0.005 (in contrast to a much lower frequency without control, well below -0.015), the control measures may be overly conservative and economically inefficient. Thus, the economic indicator is defined as the probability that the system’s nadir frequency exceeds -0.005 across 700 scenarios.

\begin{figure*}[t] %可选参数 h t b p，代表允许图片出现的位置，h表示此处附近，t表示顶部，b表示底部，p表示单独一页，H表示固定此处
    \centering
    % \vspace{-0.2cm}  %调整图片与上文的垂直距离
    % \setlength{\abovecaptionskip}{-0.05cm}   %调整图片标题与图距离
    % \setlength{\belowcaptionskip}{-2cm}   %调整图片标题与下文距离
    \includegraphics[width=16cm]{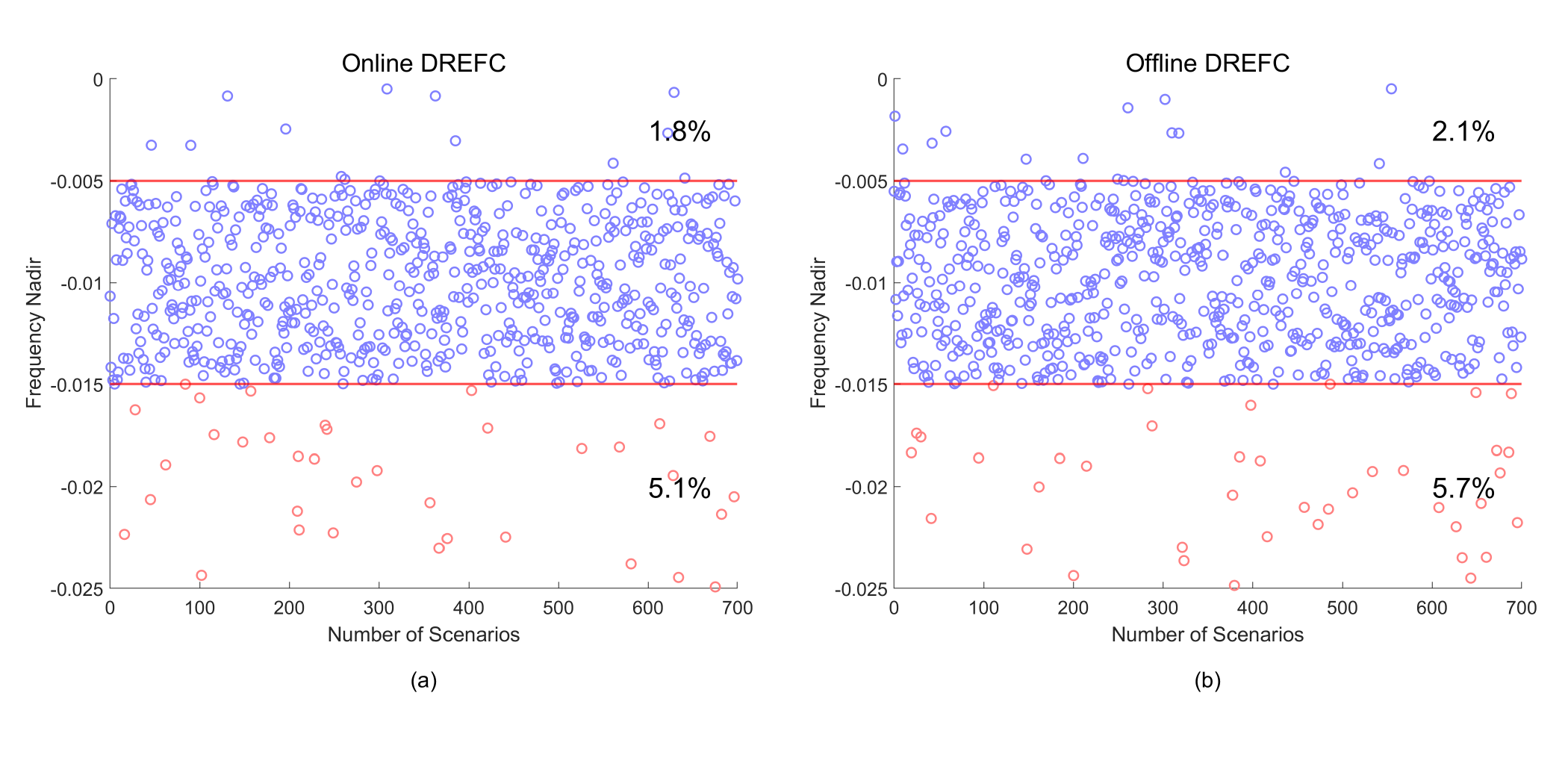}
    \caption{Frequency nadir of 700 scenarios with proposed online DREFC method when the worst distribution is used to generate the possible PEFNs.} \label{DC_regulation_combined}
\end{figure*}

Based on the definitions of the aforementioned safety and economic indicators (with both indicators being more favorable the larger their values), we compared DREFC with and without online update. The results are illustrated in Fig. \ref{DC_regulation_combined}. The results reveal that the security indicator for DREFC with online update is 94.9\%, compared to 94.7\% for DREFC without one. Similarly, the economic indicator for DREFC with online updates stands at 2\%, while it is 1.8\% for DREFC without online updates. This demonstrates that, compared to DREFC with online updates, DREFC without online updates exhibits both a lower security indicator and a lower economic indicator. 

The results above suggest that when accuracy is insufficient, certain scenarios may be erroneously estimated as overly conservative, leading to excessive DC power regulation. At the same time, other scenarios may be estimated too optimistically, resulting in insufficient DC power regulation, ultimately reducing the probability of maintaining system frequency within the safe range. Thus, compared to DREFC without online updates, DREFC with online updates can ensure system frequency safety with a higher probability while also reducing unnecessary control actions.

\subsection{Computation time}

To demonstrate the computational efficiency of DREFC, we generated 1,000 possible PEFNs from the reference distribution in Section. \ref{Modelling_accuracy_of_GMM}. These 1,000 PEFNs were then applied to Eq. \eqref{eq_2b}, creating 1,000 distinct scenarios. For each of these scenarios, the optimal control strategy was solved using both the proposed DREFC method (offline) and the traditional scenario-based SO method. 
% To ensure fairness, neither of the two methods undergo online updates. 

\begin{figure}[t] %可选参数 h t b p，代表允许图片出现的位置，h表示此处附近，t表示顶部，b表示底部，p表示单独一页，H表示固定此处
    \centering
    % \vspace{-0.2cm}  %调整图片与上文的垂直距离
    % \setlength{\abovecaptionskip}{-0.05cm}   %调整图片标题与图距离
    % \setlength{\belowcaptionskip}{-2cm}   %调整图片标题与下文距离
    \includegraphics[width=8cm]{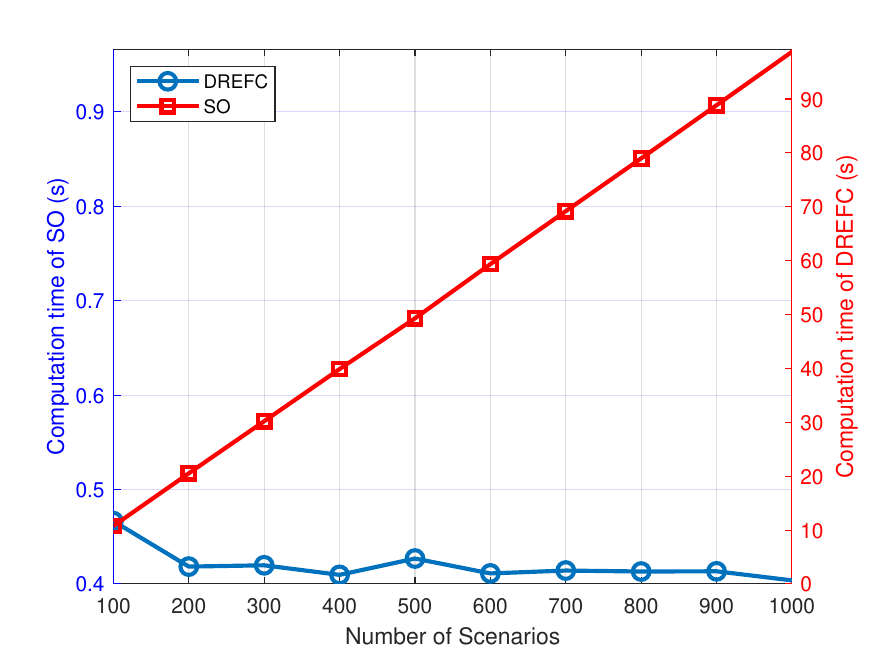}
    \caption{Computation time of proposed DREFC and traditional SO method.} \label{DRO_timing_1}
\end{figure}

The computation times were recorded and are shown in Fig. \ref{DRO_timing_1}. It is evident that as the number of scenarios increases, the computation time of the SO method rises significantly, while the DREFC method maintains a stable computation time of approximately 400 ms. This result proves that unlike the method in which the computational load increases with the amount of historical data, DREFC theoretically maintains consistent computational efficiency.

\subsection{Control Cost}
\begin{figure}[t] %可选参数 h t b p，代表允许图片出现的位置，h表示此处附近，t表示顶部，b表示底部，p表示单独一页，H表示固定此处
    \centering
    % \vspace{-0.2cm}  %调整图片与上文的垂直距离
    % \setlength{\abovecaptionskip}{-0.05cm}   %调整图片标题与图距离
    % \setlength{\belowcaptionskip}{-2cm}   %调整图片标题与下文距离
    \includegraphics[width=8cm]{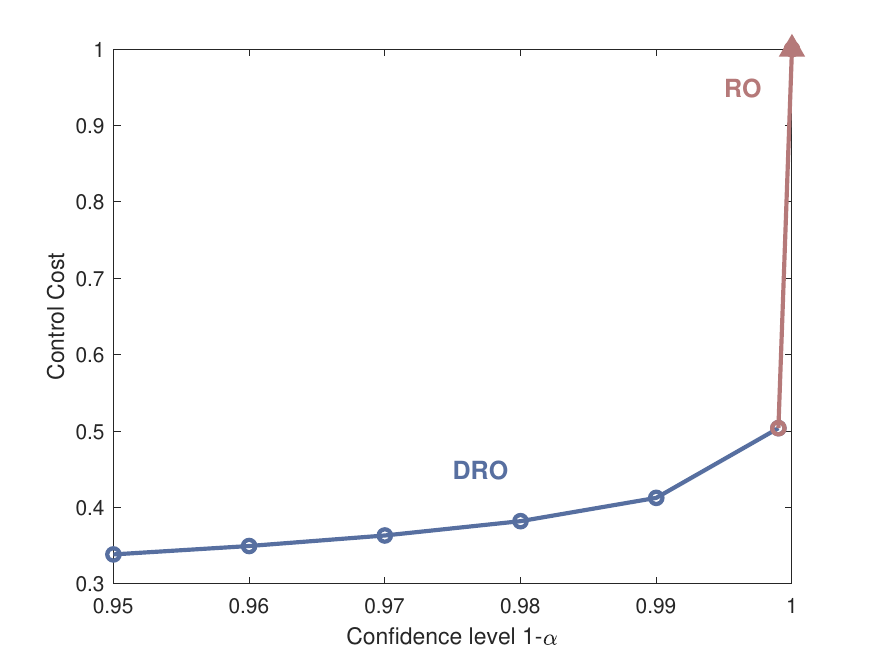}
    \caption{Control Cost of proposed DREFC method and RO method in Ref. \cite	{cao2023datadrivenfrequencyloadshedding}.} \label{Economy_1}
\end{figure}

To illustrate how DREFC can reduce the conservativeness of control and lower control costs, we calculate the control costs of DREFC under different confidence levels and compared them to the robust control method proposed in Ref. \cite{cao2023datadrivenfrequencyloadshedding}. The results are presented in Fig. \ref{Economy_1}, where the control costs of DREFC are normalized against that of the robust control method. As shown in Fig. \ref{Economy_1}, when the confidence level ranges between 0.95 and 0.99, the control costs of DREFC are only about 33-40\% of the robust control costs. Even at a confidence level of 0.999, DREFC's control costs remain approximately 50\% of those associated with robust control. This demonstrates that DREFC can significantly reduce control costs, while allowing decision-makers to adjust the model's conservativeness and robustness according to specific conditions and risk preferences.
\section{Conclusion}

This paper proposes a distributionally robust emergency frequency control model, which incorporates VaR constraints to ensure a high probability of frequency safety. With GMM-based ambiguity set described by Wassertain-type distance and approximation of VaR constraint, DREFC is reforemulated into a comptational efficient problem to meet the need for emergency control. Compared to the traditional stochastic methods for EFC, the model could be applied to the case with limited historical data. Compared with RO method, the conservativeness and robustness of the model is adjustable for decision makers according to actual situations and risk preferences. Unlike DRO based on empirical distribution, the proposed method’s computational efficiency is not significantly affected by the increase in historical data. Numerical case studies demonstrate that DREFC ensures safety, economic efficiency and low computation time in control strategies.
% \appendices
% \section{Effect of misrepresentation of Koopman eigenpairs}

% \section{}

% \section*{Appendix A}
% \vspace{-0.2cm}
% \section*{Proof of Proposition 1}

% \section*{Appendix B}
% \vspace{-0.2cm}
% \section*{Proof of Proposition 2}

\bibliographystyle{IEEEtran}
\bibliography{ref}
\end{document}